\documentstyle[epsfig, twocolumn]{esapub}
\newcommand{\OL}{$\Omega_{\Lambda}$}
\newcommand{\OO}{$\Omega$}
\newcommand{\tx}[1]{\textrm{#1}}
\newcommand{\kms}{km~$\tx{s}^{-1}$}

\newcommand{\RE}{$R_{\tx{e}}$}

\newcommand{\pa}[1]{\protect\astroncite{#1}}
\newcommand{\SBE}{SB$_{\tx{e}}$}

\newcommand{\be}{\begin{equation}}
\newcommand{\ee}{\end{equation}}

\begin{document}
\twocolumn
\setlength{\parindent}{0pt}
\setlength{\parskip}{ 10pt plus 1pt minus 1pt}
\setlength{\hoffset}{-1.5truecm}
\setlength{\textwidth}{ 17.1truecm }
\setlength{\columnsep}{1truecm }
\setlength{\columnseprule}{0pt}
\setlength{\headheight}{12pt}
\setlength{\headsep}{20pt}
\pagestyle{esapubheadings}

\title{\bf THE FUNDAMENTAL PLANE OF FIELD EARLY-TYPE GALAXIES \\ AT INTERMEDIATE REDSHIFT}
\author{Tommaso Treu$^{1,2}$, Massimo Stiavelli$^2$, Stefano Casertano$^{2,3}$, Palle M{\o}ller$^4$, Giuseppe Bertin$^1$\\
$^1$ Scuola Normale Superiore, P.za dei Cavalieri 7, I56126, Pisa,
Italy\\ 
$^2$ Space Telescope Science Institute, 3700 San Martin Dr.,
Baltimore, MD 21218, USA\\
$^3$ On assignment from the Space Science Division of the European Space Agency\\
$^4$ ESO, Karl-Schwarzschild Str. 3, D85748, Garching bei M\"unchen,
Germany\\
}
\maketitle

\begin{abstract}

We present preliminary results on the evolution of the stellar
populations of field early-type galaxies (E/S0) from $z=0.4$ to
$z=0$. The diagnostic tool used in this study is the Fundamental Plane
(FP), a tight empirical correlation between their central velocity
dispersion ($\sigma$), effective radius (\RE), and effective surface
brightness (\SBE), which is observed to hold in the local
Universe. Using HST-WFPC2 archive images and spectra obtained at the
ESO-3.6m telescope we measured the FP parameters for a sample of $\sim
30$ field E/S0s at $z=0.2-0.4$. Remarkably, field E/S0s at
intermediate redshift also define a tight FP, with scatter unchanged
with respect to that of local samples. The intermediate redshift FP is
offset from the local one, in the sense that, for given \RE\, and
$\sigma$, galaxies are brighter at $z=0.4$ than at $z=0$. The
implication of the offset of the FP in terms of passive evolution of
the stellar population depends on its star formation history. In a
single burst scenario, the stellar populations of field E/S0s were
formed at $z=0.8-1.6$ (\OO=0.3; \OL=0.7; $H_0$=50 \kms
Mpc$^{-1}$). Alternatively, the bulk of stars (90\% in mass) can be
formed at high redshift ($z\sim$ 3), and the rest in a secondary burst
occurred more recently ($z\sim0.5-0.8$).
\vspace {5pt} 

\end{abstract}

\section{Introduction}

Understanding the formation and evolution of field early-type galaxies
is a cornerstone for the entire picture of galaxy formation. In fact,
one of the predictions of hierarchical clustering models is that
early-type galaxies and their stellar populations form much later in
the field than in the core of rich clusters. This is due to the fact
that for a random Gaussian initial density field the collapse of
density peaks occurs earlier in the proximity of the large scale
overdensities that are bound to be the location of present day
clusters (Kauffmann, 1996). A number of studies have shown that the
stellar populations of early-type galaxies in the core of rich
clusters are homogeneously old (Bower, Lucey \& Ellis, 1992; Ellis et
al.\ 1997; Stanford et al.\ 1998; van Dokkum et al.\ 1998; Brown et
al.\ 2000). The few results collected so far in the field environment
indicate that the star formation history is more complex than what is
suggested by the canonical scenario of passive evolution of an old
stellar population (see e.~g. Schade et al.\ 1999; Treu \& Stiavelli,
1999, and references therein). We report here on a project that we
have just completed, aimed at gathering new information on the star
formation history of field early-type galaxies by studying the
evolution of the Fundamental Plane with redshift as a diagnostic of
stellar populations.

The Fundamental Plane (hereafter FP; Djorgovski \& Davis 1987;
Dressler et al.~1987) of early-type galaxies is defined as
\begin{equation}
\label{eq:FP} \log R_{\tx{e}} = \alpha \log~\sigma + \beta~\tx{SB}_{\tx{e}} +
\gamma,  
\end{equation} 
where $R_{\tx{e}}$ is the effective radius in kpc, $\sigma$ is the
central velocity dispersion in \kms, \SBE\, is the average surface
brightness within the effective radius in $\tx{mag arcsec}^{-2}$.  In
the following we refer, when needed, to $H_0$=h$_{50}$ 50 \kms
$\tx{Mpc}^{-1}$ and to a $\Lambda$ cosmology (\OO=0.3; \OL=0.7).  

A simple physical interpretation of the FP (Faber et al.\ 1987) can be
given by defining an effective mass,
\be 
M \equiv \frac{\sigma^2 R_{\tx{e}}}{G}, 
\ee
suggested by the Virial Theorem, by defining the luminosity in the usual way
\be
-2.5 \log L \equiv \tx{SB}_{\tx{e}} - 5 \log R_{\tx{e}} - 2.5 \log 2 \pi,
\ee
and by considering an effective mass-luminosity relation of the form
\be
L \propto M^{\eta}.
\ee
These assumptions lead directly to the FP relation given above,
provided
\be 
\alpha-10\beta+2=0, 
\label{eq:int}
\ee
(\cite{3M}), with $\eta=0.2 \alpha /\beta$. In this framework,
variations of the slopes $\alpha$, $\beta$ and the intercept $\gamma$
(typical values at $z\approx0$ are 1.25, 0.32, -8.895, in Johnson B
band, so that $\eta\approx0.8$; see Bender et al.\ 1998) as a function
of redshift are easily interpreted as general trends in the luminosity
evolution of the galactic stellar population. In particular, if we
assume fixed slopes and define for an individual galaxy (labeled by
the superscript $i$)
\be 
\gamma^i\equiv \log R_{\tx{e}}^i - \alpha \log \sigma^i - \beta \tx{SB}_{\tx{e}}^i,
\label{eq:gammai}
\ee 
the offset with respect to the prediction of the FP ($\Delta
\gamma^i \equiv \gamma^i-\gamma$) is related to the offset of the $M/L$ by
\be
\Delta \log \left( \frac {M}{L} \right)^i = -\frac{\Delta \gamma^i}{2.5 \beta}.
\ee
Similarly, the scatter (rms) of the FP can be related to the scatter
in the $M/L$ at a given $M$:
\be
rms\left(\log \frac{M}{L}\right)=\frac{rms(\gamma)}{2.5\beta}.
\ee
Therefore the very existence of the FP constrains the similarity of
the stellar populations of early-type galaxies, and it is interesting
to explore how far in the past this similarity extends.

Recently, using the Hubble Space Telescope (HST) for
photometry and ground-based spectroscopy from large telescopes, the FP
parameters of cluster E/S0s have been measured out to $z\approx0.8$
(\cite{PDdC95}; \cite{DF96}, hereafter DF96; \cite{KDFIF}, hereafter
K97; Bender et al.\ 1998; van Dokkum et al.\ 1998; Kelson et al.\
2000). The data show that the FP is well-defined up to look-back times
of $\sim 7.5$ Gyrs, indicating that the origin of the relation is hidden
at still higher redshift. The evolution of the intercept is consistent
with what is predicted by the passive evolution of an old stellar
population in a $\Lambda$ cosmology ($z_f \sim 2$; van Dokkum et al.\
1998).

In a pilot study of a sample of six early-type galaxies at
intermediate redshift (Treu et al.\ 1999; hereafter T99) we found that
the FP is also well defined in the field environment out to
$z\approx0.3$. We have now collected a larger sample of data. In this
paper we give a preliminary report of this study (the complete
data-set will be presented in Treu et al.\ 2000a, the complete
analysis in Treu et al.\ 2000b). In Section 2 we summarize the
observational results and describe the location and scatter of the
intermediate redshift field FP. In Section 3 we briefly discuss the
results in terms of passive evolution
of the stellar populations.

\section{The field FP at intermediate redshift}

\label{sec:data}

A sample of early-type galaxies was selected from the Wide Field and
Planetary Camera 2 Medium Deep Survey (Griffiths et al.\ 1994)
archival images on the basis of morphology, color, and magnitude (see
T00a for a discussion of the sample selection and data
analysis). Follow-up spectroscopy was obtained at the ESO-3.6m
telescope, yielding accurate velocity dispersions for 22 galaxies in
the sample ($z\approx0.2-0.4$).

\begin{figure}
\mbox{\epsfysize=8cm \epsfbox{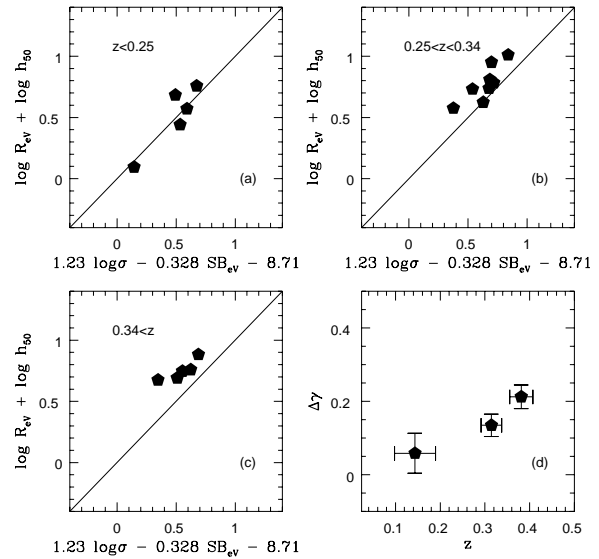}}
\caption{Fundamental Plane in V band (rest frame) at intermediate redshift. In
panels (a), (b), and (c) we plot the intermediate redshift galaxies
(pentagons) binned in redshift. The solid line represents the
FP of Coma, as measured by Lucey et al.\ (1991).  In panel (d) the
average offset of the intercept (with fixed slopes taken from Lucey et
al.\ 1991) is plotted as a function of redshift. The error bars in
panel (d) are the standard deviation of the mean. The errors in panels
(a), (b), and (c) are not shown in this edge-on projection of the FP,
since errors on the photometric parameters are correlated (see
T00a,b).}
\label{fig:FPVo0.3}
\end{figure}

\begin{figure}
\mbox{\epsfysize=8cm \epsfbox{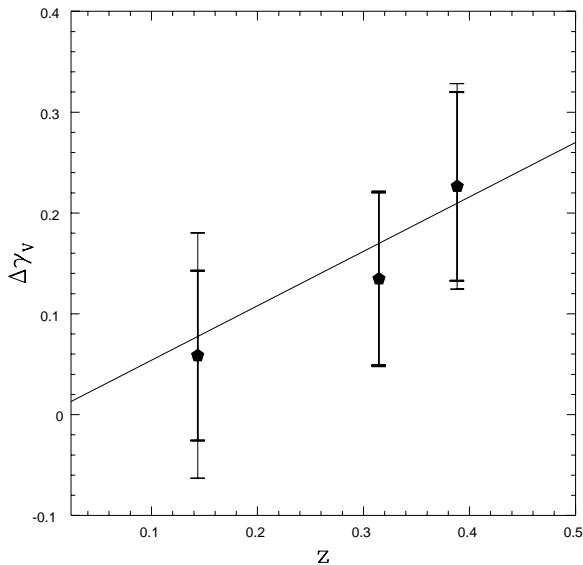}}
\caption{Evolution of the intercept of the FP and its scatter with
redshift. The solid line is a least $\chi^2$ linear fit to the
data. The thin error bars are the measured scatter, the thick error
bars are the scatter corrected for its evolutionary component (see
text).} 
\label{fig:scatz}
\end{figure}

In Figure~\ref{fig:FPVo0.3}, panels (a), (b), and (c), we plot the
location in the FP space of the intermediate redshift data points
binned in redshift. The data are in rest frame V band. The solid line
represents the FP of the Coma Cluster (Lucey et al.\ 1991). In panel
(d) we show the evolution of the intercept with redshift, with respect
to the Coma relation. Qualitatively, two main facts are evident:

\begin{enumerate}

\item At any given redshift between 0 and $\approx0.4$ the FP of {\it
field} early type galaxies is well defined. The scatter is small. No
trend of increasing scatter with redshift is noticeable within the
accuracy allowed by the small number statistics available (see
Figure~\ref{fig:scatz}).

\item The intermediate redshift data points at given \SBE\, and
$\sigma$ have larger effective radii than the value predicted by the
local relation (solid line), and hence are more luminous. The average
offset increases with redshift as shown in panel (d).

\end{enumerate}

We can quantify these statements by describing the evolution of the
intercept with a simple linear relation $\Delta\gamma = \tau z$
(Figure~\ref{fig:scatz}). A least $\chi^2$ fit yields
$\tau=0.54\pm0.02$. Taking into account selection effects (see T00b)
the 68 \% confidence interval is $0.44<\tau<0.56$.  This description
allows us to estimate the scatter of the FP in any redshift bin. In
fact, there are three contributions to the scatter: the measurement
scatter, the intrinsic scatter, and the evolutionary scatter. The
latter is simply the combined effect of the thickness of the redshift
bins and the evolution of the intercept. We can correct the measured
scatter for this projection effect by ``evolving'' linearly the
galaxies to the average redshift of the bin. The corrected scatter
obtained in this way is remarkably small and constant (0.08-0.09 in
$\gamma$, including measurement scatter  $\approx$0.05-0.06).  Given
the limited number of data points available per redshift bin the error
on the scatter is quite large ($\sim 30 $\% for a Gaussian
distribution, neglecting systematics); still a substantial
increase in the scatter can be ruled out. Larger samples are needed in
order to measure the scatter as a function of redshift accurately.

\section{Evolution of the stellar populations}

\begin{figure}
\mbox{\epsfysize=8cm \epsfbox{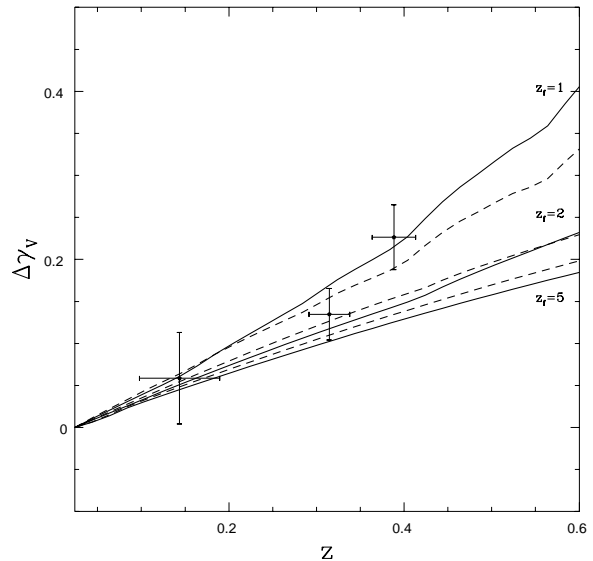}}
\caption{Single burst passive evolution models. The evolution of the
intercept of the FP is compared with the prediction of single burst
passive evolution models computed from BC96 synthetic spectra. The
solid lines represent models with Salpeter IMF (Salpeter, 1955), the
dashed lines represent models with Scalo IMF (Scalo, 1986). All models
have solar metallicity. Three redshifts of formation are assumed,
$z_f$=1,2, and 5, from top to bottom.}
\label{fig:evo}
\end{figure}

\begin{figure}
\mbox{\epsfysize=8cm \epsfbox{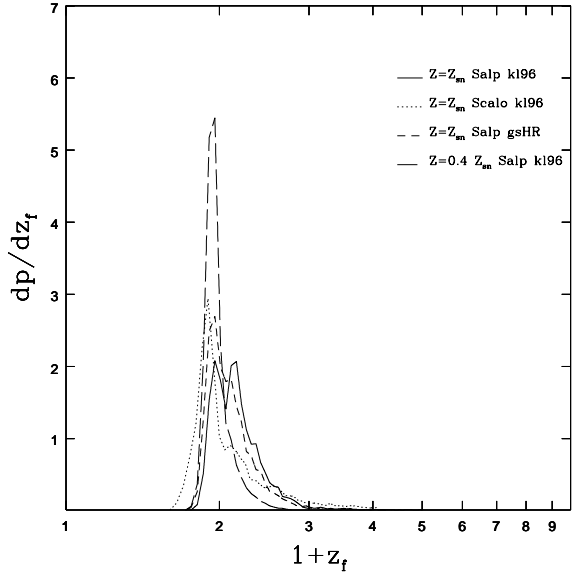}}
\caption{Single burst stellar population models. The probability
density of the redshift of formation for various realizations of the
models.  Spectral synthesis models from BC96 are used with Salpeter
and Scalo IMF, solar and 0.4 solar metallicity, atmosphere from Kurucz
or Gunn \& Striker (see Bruzual \& Charlot 1993 for
details). Independently of the details of the model, the probability
density peaks at $z_f\sim 1$.}
\label {fig:zfo0.3}
\end{figure}

In this section we compare the observed results on the FP at
intermediate redshift with the prediction of stellar population
evolution models (Bruzual \& Charlot, 1993; GISSEL96 version,
hereafter BC96). To this aim we assume that the only evolving
factor is the stellar population, that the slopes of the FP do not
change with redshift, and that the stellar mass M$_*$ is proportional
to $M$. Under these assumptions
\be
\Delta log \frac {M_*}{L} = -\frac{\Delta \gamma}{2.5 \beta},
\ee
where $\Delta$ indicates the difference of the quantity with respect
to the value found in Coma. In Figure~\ref{fig:evo} the evolution of the
FP is compared with the prediction of single burst stellar population
models formed at different formation redshifts ($z_f$).
A quantitative comparison is made
in Figure~\ref{fig:zfo0.3} where the probability density of $z_f$
given the observations is shown. The probability density is obtained
with a Bayesan-Montecarlo method described in T00b and takes into
account the selection effects. The probability peaks at $z_f\sim 1$
extending from $z_f=0.8$ to $z_f=1.6$, corresponding to present ages
of 7-10 Gyrs.

\begin{figure}
\mbox{\epsfysize=8cm \epsfbox{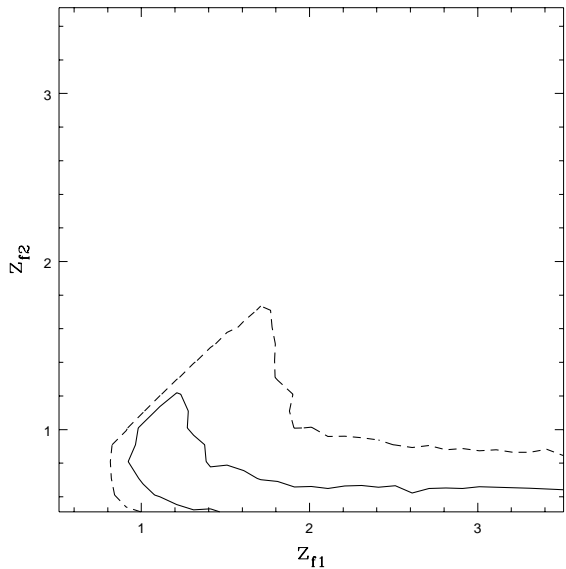}}
\caption{Model with two simple populations of stars. The older is formed at
$z_{f1}$, the younger, with a tenth of the mass, is formed at $z_{f2}$.
Contour levels of the {\it a posteriori} probability are shown as solid  
(68 \%) and dashed (95\%) lines.}
\label {fig:2Dzf}
\end{figure}

The single burst stellar population model is very useful as a
benchmark to quantify and compare different results. However, it is
clearly just a simplified picture of the star formation history of
early-type galaxies. A small amount ($<$ 10 \%) of the total mass of
moderately young stars (one Gyr or so) can alter in a significant way
the integrated colors and the $M_*/L$ of an old stellar
population. In this scenario, after a few Gyrs, the integrated colors
and the $M_*/L$ become totally indistinguishable from the ones of an old
single burst stellar population. A wealth of observational evidence
suggests that minor episodes of star formation
can occur from intermediate redshifts to the present (e.~g. Schade et
al.\ 1999; Bernardi et al.\ 1998), and it is worth addressing this
issue for our sample of field early-type galaxies. With an analysis similar to that for the
single component case we obtained (T00b) the probability density for
the redshift of formation of two stellar components: an older one
formed at $z_{f1}$ and a younger one, 10 times smaller in mass, formed
at $z_{f2}$. The contour levels of the probability density
corresponding to 68\% and 95\% probability are shown in
Figure~\ref{fig:2Dzf}. Not surprisingly, it is sufficient to have a
small mass of stars formed at $z<0.6-0.8$ to make it possible for the
rest of the stellar mass to be formed at the very beginning of the
Universe ($z_{f1}\sim$3-4, i.~e. 12-13 Gyrs ago). This scenario
is also consistent with the small scatter observed for the FP at low
and intermediate redshift, in the sense that the scatter in the FP due
to the composite ages of stellar populations is always smaller than
the observed one.

\section*{Acknowledgments}

This research has been partially funded by the Ministero
dell'Universit\`a e della Ricerca Scientifica e Tecnologica, by the
Space Telescope Science Institute (STScI) Director Discretionary Fund
grants 82216 and 82228, by the Agenzia Spaziale Italiana. It is based on
observations collected at the European Southern Observatory, La Silla,
Chile (Proposals 62.O-0592, 63.O-0468, 64.O-0281) and with the
NASA/ESA HST, obtained at the STScI, which is operated by AURA, under
NASA contract NAS5-26555.

\end{document}